\documentclass[12pt,preprint]{aastex}

\newif\ifEmulate
\def\Figure#1{\ifEmulate{#1}\fi}

\def\eq#1{\begin{equation} {#1} \end{equation}}
\let\mic=\micron
\def\E#1{\hbox{$10^{#1}$}}
\def\ga    {\hbox{$\gtrsim$}}

\def\sub#1{_{\rm #1}}

\def\about  {\hbox{$\sim$}}
\def\FB    {\hbox{$F\sub{B}$}}
\def\N     {\hbox{$\cal N$}}
\def\NT    {\hbox{${\cal N}_{\rm T}$}}
\def\Pesc  {\hbox{$P\sub{esc}$}}
\def\S     {\hbox{$S\sub{c\lambda}$}}
\def\Sd    {\hbox{$S^{\rm d}_{\rm c\lambda}$}}
\def\Si    {\hbox{$S^{\rm i}_{\rm c\lambda}$}}
\def\tV     {\hbox{$\tau\sub V$}}
\def\Ti     {\hbox{$T\sub{n\,i}$}}
\def\Tn     {\hbox{$T\sub{n}$}}
\def\Ri     {\hbox{$R\sub{i}$}}
\def\Ro     {\hbox{$R\sub{o}$}}
\def\Mo     {\hbox{$M_\odot$}}
\def\Lo     {\hbox{$L_\odot$}}
\def\lfl    {\hbox{$\lambda f\sub{\lambda}$}}
\def\deg    {\hbox{$^\circ$}}

\slugcomment{Submitted February 21, 2002}

 \shorttitle{IR from AGN}
 \shortauthors{Nenkova, Ivezi\'c \& Elitzur}

\begin{document}

\title{DUST EMISSION FROM ACTIVE GALACTIC NUCLEI}

\author{Maia Nenkova\altaffilmark{1},
       \v{Z}eljko Ivezi\'c\altaffilmark{2}
       and Moshe Elitzur\altaffilmark{1}}

\altaffiltext{1}{Department of Physics and Astronomy, University of Kentucky,
           Lexington, KY 40506--0055}
\altaffiltext{2}{Department of Astrophysical Sciences, Princeton University,
    Princeton, NJ 08544}

\email{maia@pa.uky.edu, ivezic@astro.princeton.edu, moshe@pa.uky.edu}

\begin{abstract}
Unified schemes of active galactic nuclei (AGN) require an obscuring dusty
torus around the central source, giving rise to Seyfert 1 line spectrum for
pole-on viewing and Seyfert 2 characteristics in edge-on sources. Although the
observed IR is in broad agreement with this scheme, the behavior of the 10
\mic\ silicate feature and the width of the far-IR emission peak remained
serious problems in all previous modeling efforts. We show that these problems
find a natural explanation if the dust is contained in \about\ 5--10 clouds
along radial rays through the torus. The spectral energy distributions (SED) of
both type 1 and type 2 sources are properly reproduced from different
viewpoints of the same object if the optical depth at visual of each cloud
obeys \tV\ \ga\ 60 and the clouds' mean free path increases roughly in
proportion to radial distance.
\end{abstract}


\section{INTRODUCTION}
       \label{sec:introduction}

Although there is a bewildering array of AGN classes, a unified scheme has been
emerging steadily (e.g.\ Antonucci 1993, 2002; Wills 1999).  The nuclear
activity is powered by a super\-massive (\about\E6--\E9 \Mo) black hole and its
accretion disk, which extends to \about\ 1 pc. This central engine is
surrounded by a dusty toroidal structure, extending to \ga\ 100 pc. Much of the
observed diversity is simply the result of viewing this axi\-symmetric geometry
from different angles. The torus provides anisotropic obscuration of the
central region so that sources viewed face-on are recognized as Seyfert 1
galaxies, those observed edge-on are Seyfert 2. The primary evidence for the
torus comes from spectro\-polarimetric observations of type 2 sources, which
reveal hidden type 1 emission via reflection off material situated above the
torus opening. While compelling, this evidence is only indirect in that it
involves obscuration, not direct emission by the torus itself.

An obscuring dusty torus should reradiate in IR the fraction of nuclear
luminosity it absorbs, providing direct evidence for its existence. Indeed, the
continua from most AGN show significant IR emission. Silicates, whose presence
is revealed by a spectral feature at 10 \mic, are a major constituent of
astronomical dust. Seyfert 2 sources display this feature in absorption---as
expected for edge-on viewing of an optically thick torus. However, contrary to
expectations for face-on viewing, quasars and Seyfert 1 galaxies do not show
this feature in emission, (Roche et al 1991; recent claims by Clavel et al 2000
notwithstanding). This poses a serious problem to models of the dusty torus
emission, which must suppress the silicate feature in type 1 objects while
producing it in type 2. Laor and Draine (1993) studied the effects of altering
the dust properties and conclude that suppression of the 10 \mic\ feature
requires significant silicate depletion or very large (up to 10 \mic) grains
(see also Maiolino et al 2001). This could explain type 1 SED but would require
a different explanation for type 2 objects with their prominent 10 \mic\
absorption feature.

Pier \& Krolik (1992, 1993) were the first to explore the effects of toroidal
geometry on dust radiative transfer. They note that the AGN dust must be in
clumps to protect the grains, but because of the difficulties in modeling a
clumpy distribution they approximate it with a uniform one instead. They also
neglect scattering. In spite of these approximations, their results are
encouraging. The directional dependence of their model radiation reproduces the
gross features of observed SED, indicating that the toroidal geometry captures
the essence of IR observations. Two major problems remain: the observed
emission has a far-IR peak much broader than the models can produce, and the
silicate emission feature is suppressed in face-on viewing only in a narrow,
finely-tuned range of the model parameters. Rowan-Robinson (1995) notes that
both problems could be alleviated by clumpiness. He argues that a spherical
shell is an adequate proxy for a clump, and produces type 1 IR emission from a
superposition of such concentric shells. But the shell/clump analogy is
fundamentally problematic: Since the illuminated and dark faces of a clump
produce widely different spectra, the observed SED varies with the angle
between radiation source, clump and observer (see fig.\ \ref{fig:setup}). In
contrast, outside an optically thick spherical shell the SED is always the same
irrespective of observer location because only the dark side is ever visible
(inside the shell, only the illuminated face is visible). Concentric spherical
shells cannot substitute for clumps.

We performed a more adequate calculation of IR emission from an obscuring
clumpy torus, combining realistic modeling of the emission from externally
illuminated clumps with proper handling of cloud shadowing. The latter is an
essential ingredient since the number of clouds along radial rays through the
torus must be sufficiently large to ensure that x-rays from the central engine
are virtually always attenuated in type 2 objects (Guainazzi et al 2001). Here
we report the results of our effort, which show that the SED of both type 1 and
type 2 objects are properly reproduced at different viewing angles of the same
clumpy torus---in agreement with unified schemes. In a forthcoming publication
we will provide a more detailed account of our calculations and the results.

\Figure{
 \centerline{\includegraphics[width=0.7\hsize,clip]{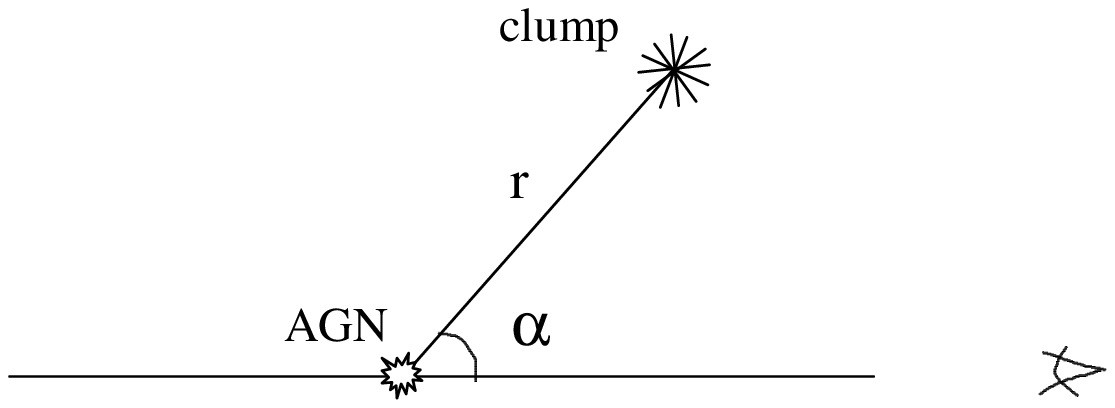}} \bigskip
 \centerline{\includegraphics[width=\hsize,clip]{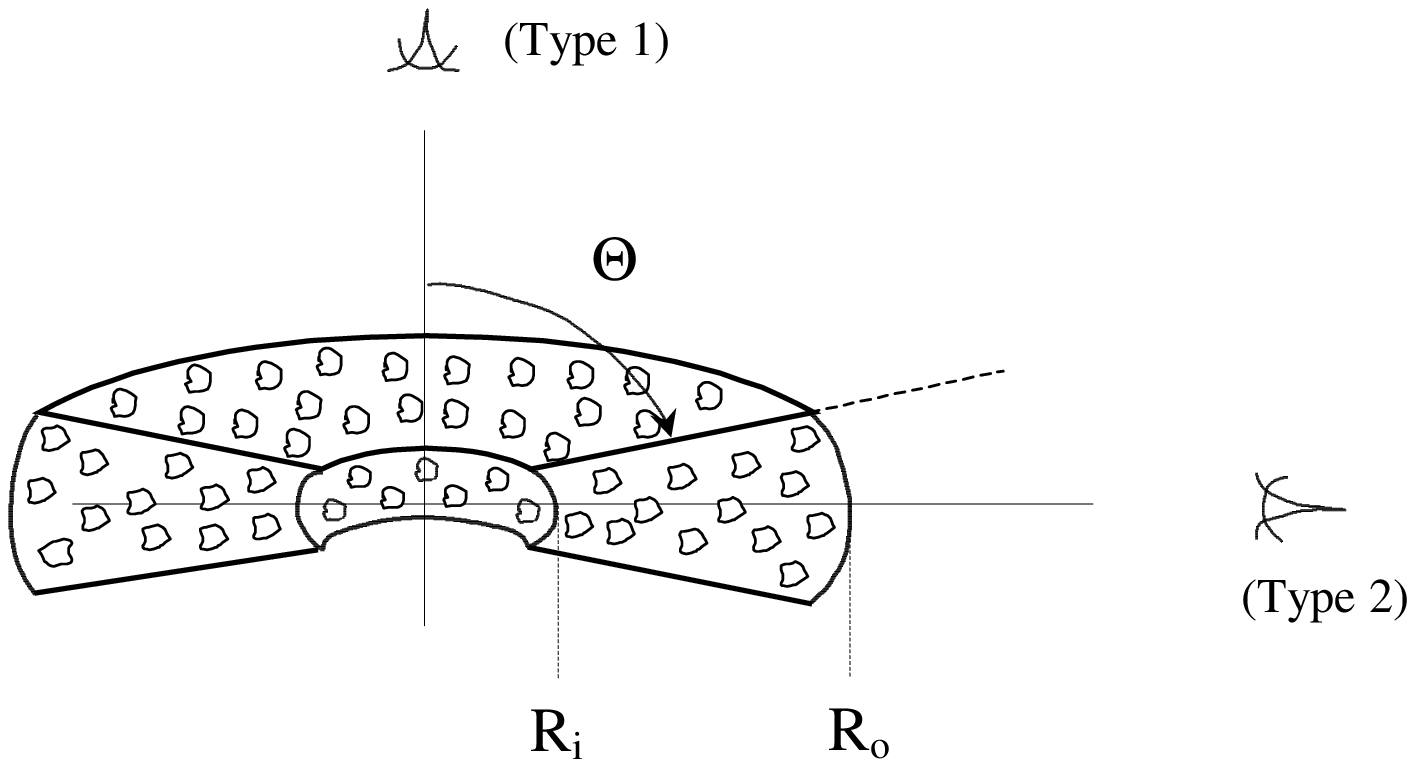}}
\figcaption[clump.ps, torus.ps]{Model geometry. Top: Positions of the AGN,
clump and observer.  As the position angle $\alpha$ varies at fixed distance
$r$, the visible fraction of the clump's illuminated area changes (like faces
of the moon) and with it the observed radiation. Bottom: The clumpy torus.
\label{fig:setup}}
}

\section{EMISSION FROM A CLUMPY MEDIUM}

Consider a medium where the dust is in clouds. For simplicity, each cloud has
the same optical depth $\tau_\lambda$.  Along a given path, the mean number of
clouds encountered in segment $ds$ is $d\N(s)= ds/\ell(s)$, where $\ell =
(n_cA_c)^{-1}$ is the mean free path between clouds ($A_c$ is the cloud area
perpendicular to the path and $n_c$ is the number density of clouds). As long
as $\ell \gg R_c$, where $R_c$ is the cloud radius, each cloud can be
considered a point source of intensity \S, and the intensity generated along
the segment is $\S d\N$.  Denote by \Pesc\ the probability that this radiation
propagate along the rest of the path without absorption by any other cloud.
Natta \& Panagia (1984) show that Poisson statistics for the distribution of
clumps yields
\eq{\label{eq:Pesc}
  \Pesc = e^{-t_\lambda},            \qquad \hbox{where}\quad
  t_\lambda  = \N(s)( 1 - e^{-\tau_\lambda} )
}
and $\N(s) = \int_sd\N$ is the mean number of clouds along the rest of the
path. The contribution of segment $ds$ to the emerging intensity is simply
$\Pesc\S d\N$ and the flux from the cloud distribution at distance $D$ is
\eq{\label{eq:FC}
  F^{\rm C}_\lambda = {1\over D^2} \int dA
        \int e^{-t_\lambda} \S(s)d\N(s)
}
where $dA$ is the surface area element perpendicular to the line of sight.
Given the geometry, the flux can be calculated from this expression once \S\ is
known.

Clouds are heated by radiation from both the AGN and all other clouds. Consider
first the direct heating by the AGN (figure \ref{fig:setup}). Since our
interest is in optically thick clouds only, the dust temperature is much higher
on the illuminated face than any other part of the surface.  Therefore the
cloud emission varies strongly with direction, and the corresponding source
function $\Sd(r,\alpha)$ depends on both distance $r$ and position angle
$\alpha$. The clump shape is of course arbitrary, and we have constructed
``synthetic clumps" by averaging the emission from an illuminated slab over all
possible slab orientations. This procedure utilizes an exact solution of
radiative transfer for external illumination and also reproduces the
$\alpha$-dependence of the observed fraction of illuminated area on the surface
of a spherical-like object.

\medskip
\Figure{
 \centerline{\includegraphics[width=\hsize,clip]{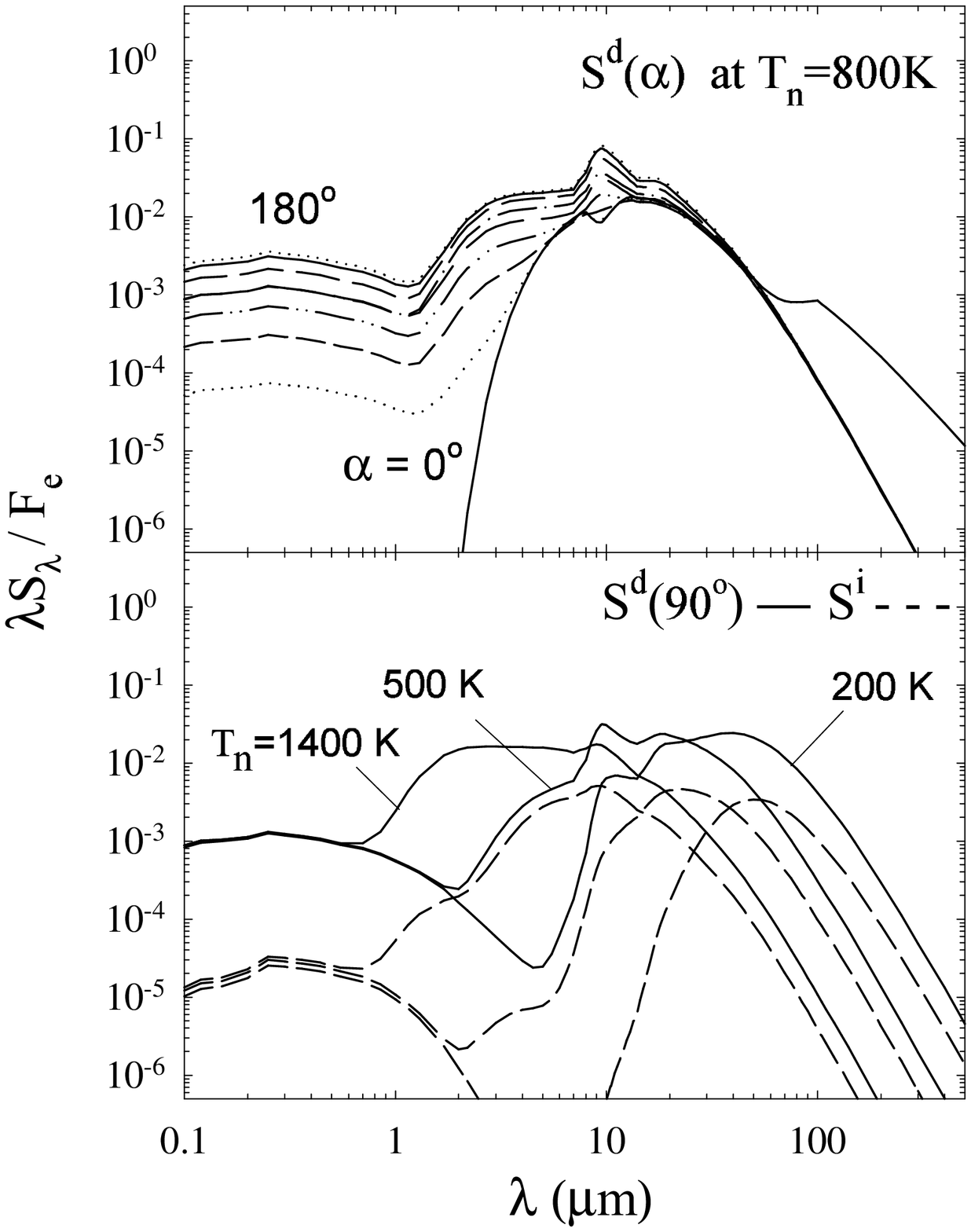}} \bigskip
\figcaption[Sd-Si.ps]{Emission from \tV\ = 100 clumps normalized to the AGN
local bolometric flux $F_e = L/4\pi r^2$. Top: The source \Sd\ for directly
heated clouds at radial distance where \Tn\ = 800 K ($r$ = 4$L_{12}^{1/2}$ pc
for this \tV). The position angle $\alpha$ is shown at 22.5\deg\ intervals.
Bottom: Emission of directly (\Sd, full lines) and indirectly (\Si, dashed
lines) heated clouds at distances where \Tn\ = 1400, 500 and 200K, or
$r/L_{12}^{1/2}$ = 1, 10 and 100 pc, respectively. The source function for
direct heating is shown at $\alpha$ = 90\deg. \label{fig:Sclump}}
}
\medskip

We conducted detailed calculations with the code DUSTY (Ivezi\'c, Nenkova \&
Elitzur 1999). The code preforms an exact solution of the slab radiative
transfer problem including dust absorption, emission and scattering. The
optical depth across the slab thickness at wavelength $\lambda$ is
$\tau_\lambda = q_\lambda\tV$ where \tV\ is the optical depth at visual and
$q_\lambda$ is the proper efficiency factor of standard interstellar dust. With
AGN luminosity $L_{12} = L/\E{12} \Lo$, the illuminating bolometric flux is
$F_e = L/4\pi r^2$. Its spectral shape, typical for AGN, is $\lfl$ = constant
in the wavelength range 0.01--0.1 \mic\ and $\propto \lambda^{-0.5}$ for
0.1--100 \mic\ (cf Pier \& Krolik 92, Laor \& Draine 93). For any slab
orientation, the solution of the radiative transfer problem determines the
temperature run in the slab and the intensity it emits toward any direction.
Therefore the dust temperature \Tn\ on the illuminated face of a slab normal to
the radius vector can replace the external flux $F_e$ as a specifier of
location.

The top panel of figure \ref{fig:Sclump} shows typical results for \Sd. The
clump spectrum was constructed by averaging over all slab orientations the
solutions for slabs with \tV\ = 100 at radial distances where \Tn\ = 800 K
(corresponding to $r$ = 4\,$L_{12}^{1/2}$ pc). The 10 \mic\ silicate feature is
seen in absorption at small $\alpha$, switching to increasingly more pronounced
emission as $\alpha$ increases and a larger fraction of the clump illuminated
face becomes visible.

An exact calculation of the effects of diffuse radiation from all clouds is a
formidable task, akin to a full solution of the radiative transfer problem in
which individual dust particles themselves are dusty clouds. However, compared
with the AGN radiation, heating by the diffuse radiation is highly inefficient
because it involves long wavelengths, and thus can be neglected. Emission from
clouds with direct view of the AGN is well described by \Sd. But clouds whose
line-of-sight to the AGN is blocked by another cloud will be heated only
indirectly by the diffuse radiation. We approximate the diffuse radiation field
at $r$ by angle-averaging over $\alpha$ the emission $\Sd(r,\alpha)$ from
clouds directly heated by the AGN.  In this radiation field we imbed a sphere
with constant density profile and solve for its temperature and emission. Our
approximation for the source function of indirectly illuminated clouds is $\Si
= F/\Omega$, where $F$ is the flux and $\Omega$ the solid angle of such a
sphere at a large distance.  The bottom panel of figure \ref{fig:Sclump} shows
a sample of \Si, together with the corresponding source functions for direct
heating.

At distance $r$ the probability for unhindered view of the AGN is $p(r) =
e^{-{\cal N}(r)}$, where $\N(r) = \int^r dr/\ell$ is the mean number of clouds
to the AGN. Our approximation for the clump source function is
\eq{\label{eq:S}
            \S = p\,\Sd + (1 - p)\,\Si\,.
}
The steps we outlined can be iterated to become an exact solution scheme for
radiative transfer in clumpy medium. The small magnitude at short wavelengths,
which control heating, of \Si/\Sd\ evident in figure \ref{fig:Sclump} indicates
that this process should converge rapidly. Eq.\ \ref{eq:S} is equivalent to its
first two steps.

\section{MODEL RESULTS}

We model the AGN obscuring region as a toroidal distribution of dusty clouds,
shown in figure \ref{fig:setup} . Instead of the inner radius \Ri\ we specify
as input \Ti\ = \Tn(\Ri), the surface temperature of a normally illuminated
slab at \Ri. In all calculations we set \Ti\ = 1400 K, the temperature below
which both silicate and graphite grains exist; this choice implies \Ri\ =
$1.2\,L_{12}^{1/2}$ pc for a \tV\ = 100 cloud. The geometry requires two
additional input parameters: The torus thickness $Y = \Ro/\Ri$, equivalent to
the lowest \Tn\ in the torus, and $\Theta$, the conical opening half-angle.
Assuming power law variation for the mean free path $\ell \propto r^q$, the
cloud distribution is described by the two input parameters $q$ and $\NT =
\int_{R_{\rm i}}^{R_{\rm o}}dr/\ell$, the number of clouds along a radial ray
through the torus. The final free parameter is \tV, the optical depth of each
cloud. The grain efficiency factor $q_\lambda$ and the AGN spectral shape
$f_\lambda$ are set to their standard values. The observed flux is determined
from
\eq{
    F_\lambda = {L\over4\pi D^2}\,f_\lambda e^{-t_{\lambda\rm T}}
              +  F^{\rm C}_\lambda
}
where $F^{\rm C}_\lambda$ is calculated from eq.\ \ref{eq:FC} and
$t_{\lambda\rm T}$ from eq.\ \ref{eq:Pesc} with $\N(s) = \NT$ for type 2
sources and zero for type 1.  Extracting the overall bolometric flux $\FB =
L/4\pi D^2$, the SED is controlled by \Ti\ and five dimensionless free
parameters: $Y$ and $\Theta$ specify the geometry, \tV\ the individual clouds
and \NT\ and $q$ the cloud distribution.

\medskip
\Figure{ \centerline{\includegraphics[width=\hsize,clip]{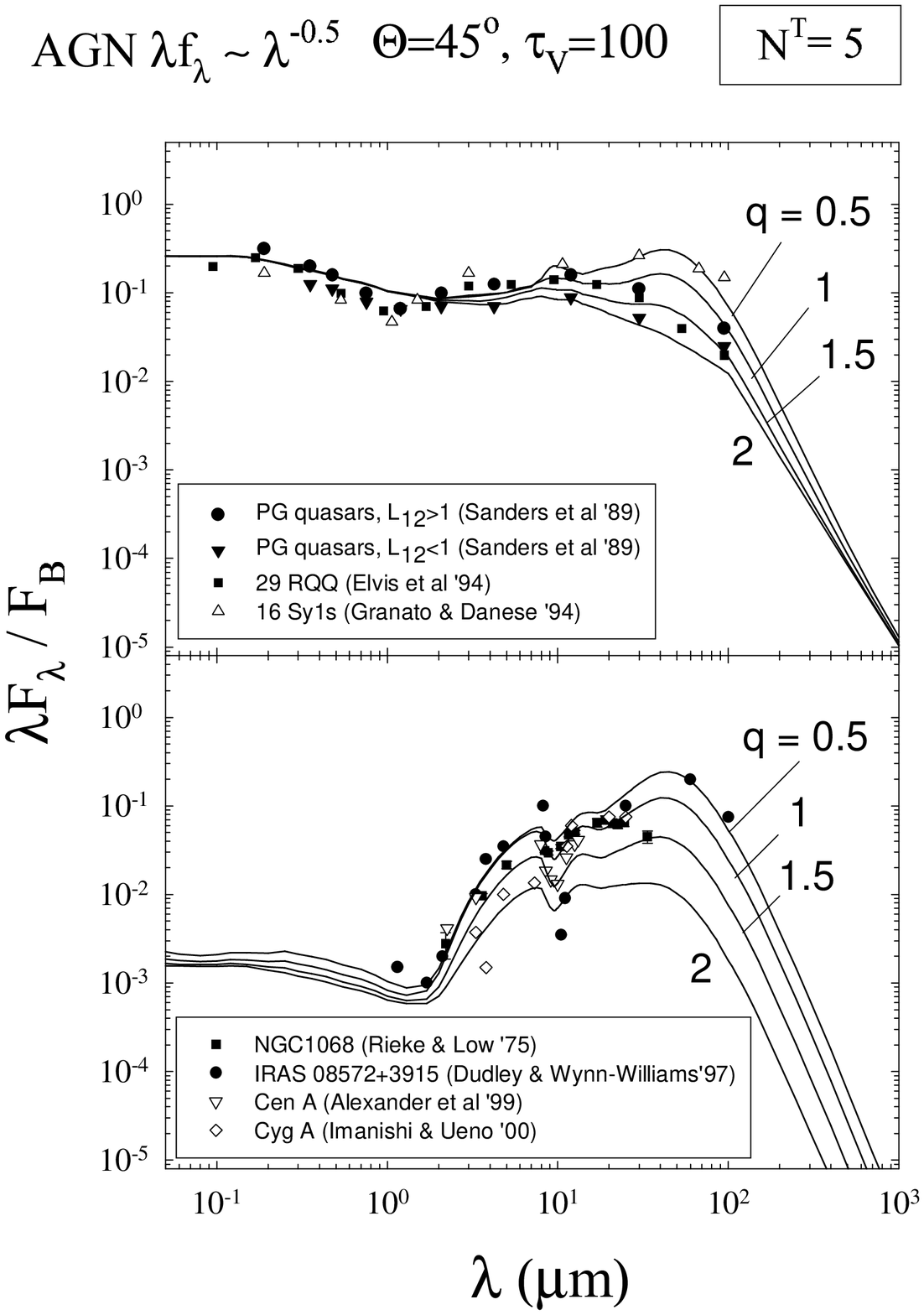}}
\figcaption[Fig-fits.ps]{Modeling and observations of type 1 (top panel) and
type 2 (bottom) sources. Lines are model results for axial and equatorial views
of the clumpy torus shown in figure 1, with $\FB = L/4\pi D^2$.  Radial rays
through the torus have 5 clouds on average, each with \tV\ = 100. The clouds'
mean free path varies as $r^q$, with the indicated $q$. The torus inner radius
corresponds to temperature \Ti\ = 1400 K (\Ri\ = $1.2\,L_{12}^{1/2}$ pc) and it
has \Ro\ = 100\Ri\ and $\Theta$ = 45\deg. Data for type 1 are average spectra
from the indicated compilations, type 2 are for individual objects. Each data
set was scaled for rough matching with the models.
     \label{fig:SED}}
}
\medskip

We performed extensive calculations and find a large range of the free
parameters that yield satisfactory agreement with observations. A detailed
report of our results will be provided elsewhere, here we show in figure
\ref{fig:SED} a representative model.  Our aim is to reproduce the typical SED
of the type 1 and type 2 families, not to fit in detail any particular object.
Data points for type 1 sources show average spectra for radio-quiet quasars
(Sanders et al 1989, Elvis et al 1994) and Seyfert 1 galaxies (Granato \&
Danese 1994). Because of the high obscuration of the AGN in type 2 sources,
fluxes for their nuclear regions properly extracted out of the contributions of
the host galaxy and starburst regions are scarce. We show data for reliable
measurements of four individual objects. The data were scaled for a rough match
of the model results without attempting a best fit. Detailed fitting of type 2
objects, when warranted, will determine their bolometric flux \FB, a quantity
inaccessible for direct measurement because the bulk of the flux is emitted
away from Earth.

Our models reproduce the broad IR bump extending to \about\ 100 \mic, as
observed. Furthermore, the 10 \mic\ feature appears in absorption in equatorial
viewing but is smeared out in axial viewing in spite of its prominence in
emission from directly-heated individual clouds (fig.\ \ref{fig:Sclump}). The
reason is that most of these clouds are blocked from view even along the axis.
We find satisfactory results for conical opening $\Theta = 45\pm15 \deg$. The
torus thickness can vary in the range $Y$ = 50--250, i.e., \Ro/$L_{12}^{1/2}$ =
60--300 pc. At smaller $Y$ the IR bump is not broad enough, larger $Y$ produce
too much 20--100\mic\ emission. The number of clouds is \NT\ = 5--10. When $\NT
< 5$ the 10\mic\ feature appears in emission in pole-on spectra, $\NT > 10$
shifts the spectrum to longer wavelengths resulting in too little 1--10\mic\
emission and too much 20--100\mic.  The power index should be $q = 1\pm0.5$. At
larger $q$ the far-IR bump is not high enough, as is evident in figure
\ref{fig:SED}, at smaller $q$ it is too high.

Significantly, the only constraint on the optical depth of individual clouds is
\tV\ \ga\ 60. The results vary only slightly when \tV\ increases from 60 to
100, and hardly at all during further increase. The reason is simple. The
dependence on \tV\ arises from the probability for photon escape and the cloud
source function. From eq.\ \ref{eq:Pesc}, $P\sub{esc} = e^{-\cal N}$ whenever
$\tau_\lambda \gg 1$, and at \tV\ \ga\ 60 this condition is met at all relevant
wavelengths. Similarly, because the clouds are heated externally, only their
surface is heated significantly when \tV\ is large. Increasing \tV\ further
only adds cool material, thus \S\ saturates for all relevant $\lambda$, similar
to standard black-body emission. Extending our calculations all the way to \tV\
= 500, we have verified that increasing \tV\ indeed has no effect on the model
results.

\section{DISCUSSION}

The cloud distribution is described by $q$ and \NT, individual clouds by the
optical depth \tV. No other cloud property was specified. The cloud size enters
only indirectly through the underlying assumption $R_c \ll \ell$. If $V_c
\simeq A_cR_c$ is the volume of a single cloud, the volume filling factor of
the cloud population is $\phi = n_cV_c = R_c/\ell$. Our calculations apply to
small filling factors, $\phi \ll 1$, the Pier \& Krolik model involves the
opposite limit $\phi = 1$.  The mean free path scale can be determined from
$\ell(\Ri)/\Ri = (1/\NT)\int_1^Y y^{-q}dy$. The model with $q$ = 1, \NT\ = 5
and $Y$ = 100 has $\ell(\Ri) = 0.9\Ri\ = 1.1L_{12}^{1/2}$ pc. A reasonable
realization of this model, though not unique, is a constant $\phi = 0.1$
throughout the torus so that $R_c$ = 0.09r. When \tV\ = 100 and $L_{12}$ = 1,
the clouds vary from $R_c$ \about\ 0.1 pc with gas density  \about\ 3\,\E5\
cm$^{-3}$ in the torus inner regions to $R_c$ \about\ 10 pc with density
\about\ 3\,\E3\ cm$^{-3}$ at the outer edge. The torus could contain additional
clouds with smaller \tV\ without significantly affecting the SED.

The properties we deduce for the torus agree with independent estimates. Based
on 88 Seyfert galaxies Schmitt et al (2001) conclude that $\Theta$ = 48\deg, at
the center of the range we find. Our minimal cloud number \NT\ = 5 implies a
probability $\le e^{-5}$ = 7\,\E{-3} for unattenuated view of the AGN in type 2
sources, in agreement with the findings of Guainazzi et al (2001). From x-ray
measurements of 73 Seyfert 2 galaxies Bassani et al (1999) find a large
variation in column density, with a mean of $3\,\E{23}$ cm$^{-2}$ and \ga\
\E{24} cm$^{-2}$ for as many as a third of the sources. The mean is comparable
with the column density of our torus for 5 clouds of \tV\ \ga\ 60 each and
standard dust-to-gas ratio. Furthermore, the similarity of SED among type 2
sources in spite of the large x-rays column variation is a natural consequence
of our model: \tV\ is bounded only from below, the SED remains the same at all
$\lambda\ \ga\ 1\mic$ for arbitrary increase in overall column. But x-ray
absorption does vary with the latter because the optical depth for Thompson
scattering is only \about\ \E{-2}\tV. Sources with small columns may show up in
x-ray absorption while selectively excluded from IR observations because of
their weak emission.

Our results add strong support for unification schemes of AGN. In accordance
with such schemes, the IR emission from both type 1 and type 2 sources is
reproduced at different viewing of the same geometry when proper account is
taken of the torus clumpiness.

\acknowledgements

We have greatly benefitted from discussions with many colleagues, especially
J.\ Conway, A.\ Laor, N.\ Levenson and H.\ Netzer. Support by NASA and NSF is
gratefully acknowledged.

\ifEmulate\end{document}\fi


\newpage

{
 \centerline{\includegraphics[width=0.7\hsize,clip]{clump.ps}} \bigskip
 \centerline{\includegraphics[width=\hsize,clip]{torus.ps}}
\figcaption[clump.ps, torus.ps]{Model geometry. Top: Positions of the AGN,
clump and observer.  As the position angle $\alpha$ varies at fixed distance
$r$, the visible fraction of the clump's illuminated area changes and with it
the observed radiation. Bottom: The clumpy torus. \label{fig:setup}}
}

\newpage

{
 \centerline{\includegraphics[width=0.75\hsize,clip]{Sd-Si.ps}} \bigskip
\figcaption[Sd-Si.ps]{Emission from \tV\ = 100 clumps normalized to the AGN
local bolometric flux $F_e = L/4\pi r^2$. Top: The source \Sd\ for directly
heated clouds at distance where \Tn\ = 800 K ($r$ = 4$L_{12}^{1/2}$ pc for this
\tV). The position angle $\alpha$ is shown at 22.5\deg\ intervals. Bottom:
Emission of directly (\Sd, full lines) and indirectly (\Si, dashed lines)
heated clouds at distances where \Tn\ = 1400, 500 and 200K, or $r/L_{12}^{1/2}$
= 1, 10, 100 pc, respectively. The source function for direct heating is shown
at $\alpha$ = 90. \label{fig:Sclump}}
}

\newpage

{ \centerline{\includegraphics[width=0.78\hsize,clip]{Fig-fits.ps}}
\figcaption[Fig-fits.ps]{Modeling and observations of type 1 (top panel) and
type 2 (bottom) sources. Lines are model results for the axial and equatorial
views of the clumpy torus shown in figure 1, with $\FB = L/4\pi D^2$.  Radial
rays through the torus have 5 clouds on average, each with \tV\ = 100. The
clouds' mean free path varies as $r^q$, with the indicated $q$. The torus inner
radius corresponds to temperature \Ti\ = 1400 K (\Ri\ = $1.2\,L_{12}^{1/2}$ pc)
and it has \Ro\ = 100\Ri\ and $\Theta$ = 45\deg. Data for type 1 are composite
spectra from the indicated compilations, type 2 are for individual objects.
Each data set was scaled for rough matching with the models.
     \label{fig:SED}}
}

\end{document}